\documentclass[%
 reprint,
 amsmath,amssymb,
 aps,
longbibliography]{revtex4-1}

\usepackage{graphicx}
\usepackage{dcolumn}

\makeatletter

\newcommand{\Rmnum}[1]{\expandafter\@slowromancap\romannumeral #1@}
\makeatother

\begin{document}

\preprint{APS/123-QED}

\title{Rydberg state ionization dynamics and tunnel ionization rates in strong electric fields}

\author{K. Gawlas}
\author{S. D. Hogan}%
\affiliation{%
Department of Physics and Astronomy, University College London, Gower Street, London WC1E 6BT, U.K.
}%

\date{\today}

\begin{abstract}
Tunnel ionization rates of triplet Rydberg states in helium with principal quantum numbers close to 37 have been measured in electric fields at the classical ionization threshold of $\sim$ 197~V/cm. The measurements were performed in the time domain by combining high-resolution continuous-wave laser photoexcitation and pulsed electric field ionization. The observed tunnel ionization rates range from 10$^5$ s$^{-1}$ to 10$^7$ s$^{-1}$ and have, together with the measured atomic energy-level structure in the corresponding electric fields, been compared to the results of calculations of the eigenvalues of the Hamiltonian matrix describing the atoms in the presence of the fields to which complex absorbing potentials have been introduced. The comparison of the measured tunnel ionization rates with the results of these, and additional calculations for hydrogen-like Rydberg states performed using semi-empirical methods, have allowed the accuracy of these methods of calculation to be tested. For the particular eigenstates studied the measured ionization rates are $\sim~5$ times larger than those obtained from semi-empirical expressions. 

\end{abstract}

\pacs{Valid PACS appear here}
\maketitle


\begin{figure*}
	   \includegraphics[width=0.8\linewidth]{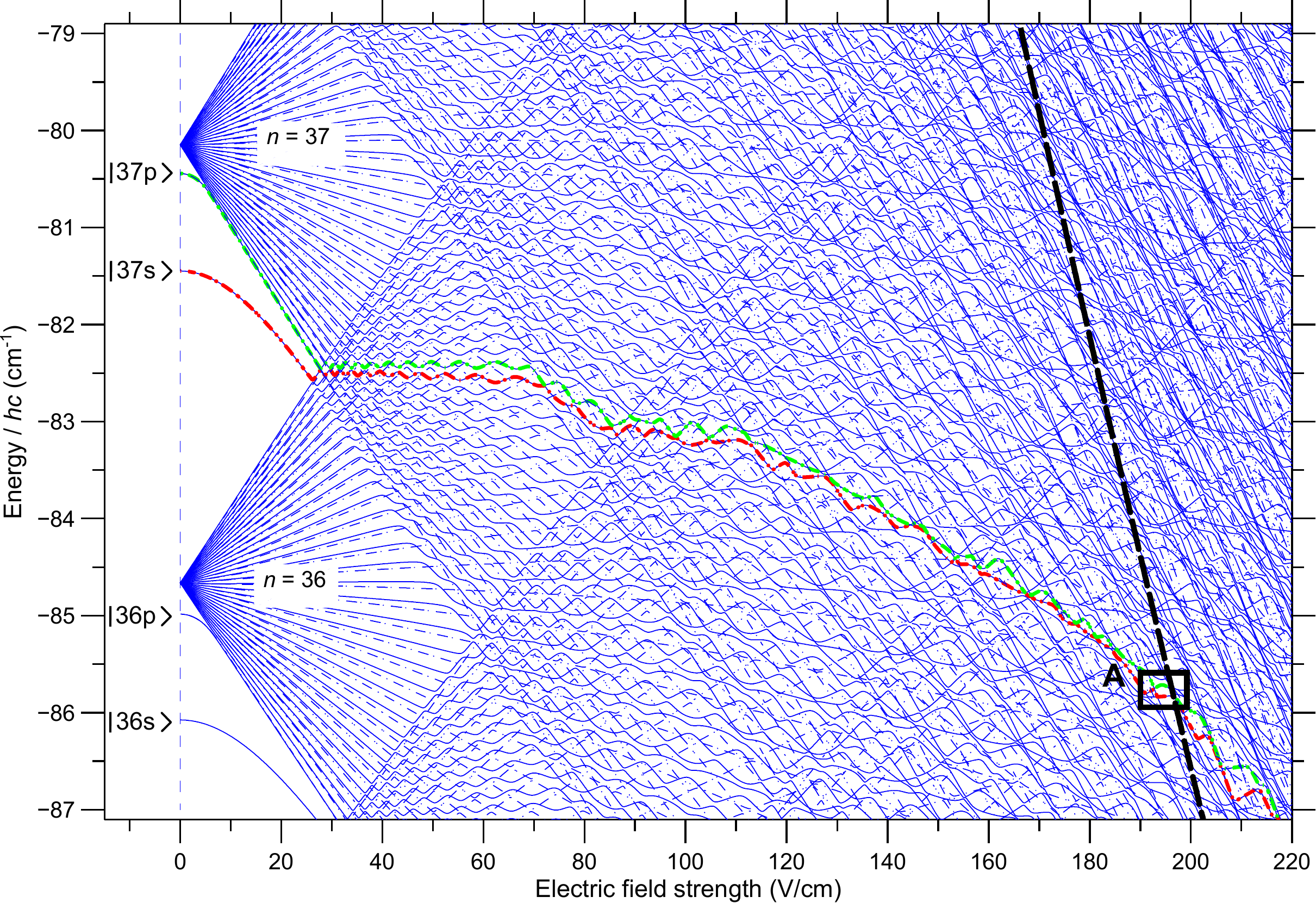}
\caption[width=2\linewidth]{Calculated Stark map of triplet Rydberg states in helium in the vicinity of the field-free $n$ = 36 and $n$ = 37 states. Sub-levels for which $m$ = 0 ($|m|$ = 1) are indicated by the continuous (dash-dotted) curves. The classical ionization threshold is denoted by the long-dashed black curve between 166 and 202 V/cm. The states that evolve adiabatically from the $\big|$37s$\big\rangle$ and $\big|$37p$\big\rangle$ field-free Rydberg states as the field is increased are indicated by the red and green thick dash-dotted curves, respectively. The region enclosed by the rectangle labeled A is discussed in the text.}\label{fig1}
\end{figure*}

\section{\label{sec:level1}Introduction}
Rydberg states of atoms and molecules represent ideal model systems with which to perform detailed studies of electric field ionization, and develop schemes with which to manipulate and control ionization processes~\cite{Feynman.2015}. Precise measurements of ionization dynamics and electric-field ionization rates of Rydberg states are of interest in the optimization of schemes for quantum-state-selective detection~\cite{Ducas.1975}. These schemes are used  in precision microwave and millimeter wave spectroscopy~\cite{Osterwalder.1999,Deller.2018}, and in studies of collisions and interactions of Rydberg atoms and molecules in which state-changing occurs \cite{Kellert.1980,Gallagher.1992}, including, e.g., studies of resonant energy transfer in collisions of Rydberg atoms with polar ground state molecules \cite{Zhelyazkova.2017,Zhelyazkova.2017b}. The accurate characterization of these ionization processes is also of interest for studies of charge-transfer in Rydberg-atom and Rydberg-molecule$-$surface scattering \cite{Hill.2000,Lloyd.2005,Gibbard.2015}, and for the generation of high-brightness, highly monochromatic electron beams for use in high-resolution electron energy loss spectroscopy \cite{Kime.2013,McCulloch.2017}. In addition, measurements of Rydberg state tunnel ionization rates can be used in general to test and validate theoretical approaches to treating atomic and molecular ionization in strong electric fields, see, e.g.,~Ref. \cite{Svensmark.2016}. Such theoretical methods are important in the analysis and interpretation of photoelectron spectra, and spectra of high harmonics resulting from the interaction of atoms or molecules with strong laser fields~\cite{Stolow.2004,Austin.2016}. \par

In this article we report experimental and theoretical studies of tunnel ionization rates of triplet Rydberg states in helium with principal quantum numbers, $n$, close to 37. This work has been carried out in electric fields between 190 and 200 V/cm, which are in the vicinity of the classical adiabatic ionization electric field of the 1s37s $^3$S$_1$ level at $\sim$ 197 V/cm. The Rydberg states investigated have been selected for study as they are of particular interest in the refinement and optimization of detection schemes used for studies of F{\"o}rster resonance energy transfer in collisions with ground state ammonia molecules~\cite{Zhelyazkova.2017,Zhelyazkova.2017b}. The experimental results have been compared with the results of calculations performed using semi-empirical analytical methods for the treatment of hydrogen-like Rydberg states with similar characteristics to those of the states investigated, and numerical methods based on the use of complex absorbing potentials (CAPs). \par

In previous studies of electric field ionization of Rydberg states in helium it was observed that even though the low-$\ell$ states studied ($\ell$ is the orbital angular momentum quantum number) fully ionized in fields slightly higher than the classical ionization threshold, there were many oscillations in the ionization signal as the electric field strength was varied as a result of mixing between stable (bound) and unstable (ionizing) eigenstates~\cite{WillemvandeWater.1984}. Such oscillations in the ionization signal indicate that the ionization rates must vary accordingly. One way to calculate the ionization rates, to provide a more rigorous quantitative interpretation of these observations, and which was described recently in Ref. \cite{Grimmel.2017}, involves introducing a CAP to the Hamiltonian \cite{Sahoo.2000,Santra.2006} describing the atom in the presence of the field. This transforms the Hamiltonian operator associated with the time-dependent tunnel ionization problem into a non-Hermitian time-independent Hamiltonian \cite{Fring.2017}. When employed in solving the Schr{\"o}dinger equation for the system this can yield complex eigenvalues. The real parts of these eigenvalues determine the eigenenergies, while the imaginary parts represent the tunneling rates. Experimentally determined ionization rates have been inferred previously from the widths of resonances in laser spectra recorded close to threshold ionization fields \cite{Grimmel.2017,McCulloch.2017} and agreed with the results of calculations using these methods. Here we measure tunnel ionization rates, which range over almost two orders of magnitude, directly in the time domain. Over this larger dynamic range discrepancies between the experimental and calculated data are identified. \par

This article is structured as follows: In Sec.~\Rmnum{2} the theoretical methods used in the numerical calculation of the energy-level structure of the triplet Rydberg states in helium, and the calculation of tunnel ionization rates is described. The experimental setup and procedures employed to study the ionization dynamics and measure tunnel ionization rates are then presented in Sec.~\Rmnum{3}. This is followed in Sec.~\Rmnum{4} by the experimental results, where the measured spectra and ionization rates are compared to the results of the calculations. Finally, in Sec.~\Rmnum{5} conclusions are drawn. 

\section{\label{sec:level2}Theoretical background}
In an electric field, $\vec{F}=(0,0,F_{z})$, the Hamiltonian for a helium atom with a single excited Rydberg electron, neglecting spin-orbit effects, can be expressed in atomic units as \cite{Damburg.1979}
\begin{eqnarray}
\hat{H} & = \hat{H}_0 + \vec{f} \cdot \hat{\vec{r}} \\
& = \hat{H}_0 + f_z \hat{z}
\end{eqnarray}
where $\hat{H}_{0}$ is the Hamiltonian in the absence of the field, $\hat{\vec{r}}=(\hat{x},\hat{y},\hat{z})$ is the position operator, and $\vec{f}=\vec{F}/F_{0}$, where $F_{0} = 2hc R_{\mathrm{He}}/e a_{0_{\mathrm{He}}}$ with $R_{\mathrm{He}}$ and $a_{0_{\mathrm{He}}}$ the Rydberg constant and Bohr radius, respectively, corrected for the reduced mass of helium, $h$ is the Planck constant, $c$ is the speed of light in vacuum, and $e$ is the elementary charge.\par

The energy-level structure of the Rydberg states in helium can be calculated numerically by determining the eigenvalues of the Hamiltonian matrix, $\hat{H}$, in Eq. (2), for a range of electric fields of interest. In doing this in the $\big|n\ell m\big\rangle$ basis ($m$ is the azimuthal quantum number), the diagonal matrix elements associated with $\hat{H}_0$ are obtained from the Rydberg formula including the quantum defects, $\delta_{n\ell}$, for the low-$\ell$ states from~Ref. \cite{Drake.1999}. The values of these quantum defects for states with $n$~=~36 and~37 are listed in Table~1. \par
\begin{table}
\begin{tabular*}{\linewidth}{@{\extracolsep{\fill}} c c c}
\hline
$\ell$ & $\delta_{36\ell}$ & $\delta_{37\ell}$ \\
\hline
0 & 0.296\,687 & 0.296\,685 \\
1 & 0.068\,346 & 0.068\,347 \\
2 & 0.002\,886 & 0.002\,887 \\
3 & 0.000\,446 & 0.000\,446 \\
4 & 0.000\,127 & 0.000\,127 \\
\hline
\end{tabular*}
\caption{Quantum defects of $n$ = 36 and 37 triplet Rydberg states in helium \cite{Drake.1999}.}
\end{table}
The radial integrals required in the calculation of the off-diagonal matrix elements associated with the electric field term in Eq.~(2) were determined using the Numerov method. This was implemented as described in Ref.~\cite{MyronL.Zimmerman.1979} with logarithmic rescaling of the spatial coordinate, an integration stepsize of 0.005, and a pure $1/r$ potential. The corresponding angular integrals were determined using standard angular momentum algebra. \par

An example of a Stark map calculated using this approach for values of $n$ close to 37, azimuthal quantum numbers $|m|$ = 0 and~1, and fields up to 220 V/cm is displayed in Fig.~\ref{fig1}. To achieve convergence of the calculated energy-level structure to $\sim 0.001$~cm$^{-1}$ in the highest fields in this figure, a basis of states containing values of $n$ from 28 to 55, and all corresponding values of $\ell$ was required. In this figure the energy associated with the saddle point in the potential experienced by the Rydberg electron in the presence of the field, i.e., the Stark saddle point, is indicated by the thick dashed curve between 166 and 202 V/cm. The corresponding classical ionization field can be expressed as $f_{\mathrm{class}} = 1/16n^{*4}$, where $n^*$ is an effective principal quantum number. In general, for eigenstates that lie below the energy of this saddle point in a particular field, the Rydberg electron remains bound to the He$^{+}$ ion core, while states that lie above the saddle point ionize. However, for states that lie close to the Stark saddle point tunneling processes play an important role. For these states, their stabilization and ionization rates depend strongly on the characteristics of their electron charge distributions, and hence their static electric dipole moments. For example, in a given field, states that exhibit negative (positive) Stark energy shifts possess static electric dipole moments oriented parallel (anti-parallel) to the field. Consequently, the corresponding electron charge density is located predominantly on the side of the ion core close to (far from) the saddle point. As a result, in fields close to $f_{\mathrm{class}}$ states with negative Stark energy shifts readily tunnel ionize. On the other hand, in similar fields, states with positive Stark shifts have a lower probability of ionizing and must be de-polarized by the field before ionization occurs. This depolarization can occur following adiabatic traversal through an avoided crossing or by enhanced $n$-mixing in higher fields.\par

For hydrogenic Rydberg states tunnel ionization rates in strong electric fields can be calculated using the semi-empirical methods described in~Refs. \cite{Damburg.1979,Damburg.1983}. To treat these electric field ionization processes for non-hydrogenic states, such as the low-$|m|$ Rydberg states in helium that are of interest here, numerical methods must be employed. This can be achieved in a manner that is compatible with the matrix methods used in the calculation of the energy-level structure in Fig.~\ref{fig1} by the introduction of complex absorbing potentials to the expression in Eq. (2). This approach, and the methods implemented here, are based on the recent work in Ref. \cite{Grimmel.2017} in which ionization processes in Rydberg states of rubidium were studied experimentally and theoretically.\par

\begin{figure}
	   \includegraphics[width=1\linewidth]{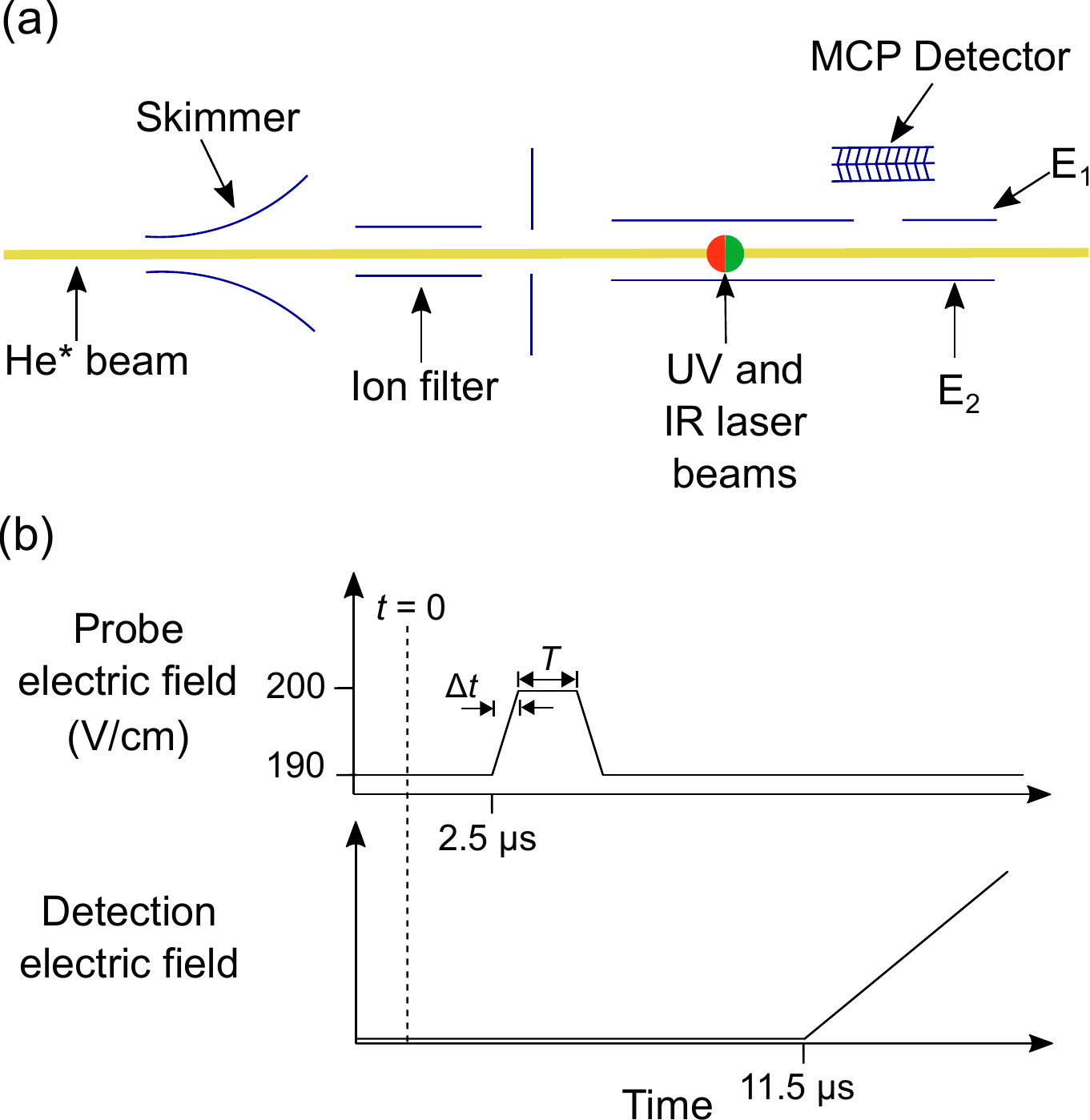}
\caption{(a) Schematic diagram of the experimental setup. (b) Timing sequence indicating the excitation (time $t=0$), electric field switching, and detection steps of the experiments.}\label{fig2}
\end{figure}

\begin{figure*}
	   \includegraphics[width=0.8\linewidth]{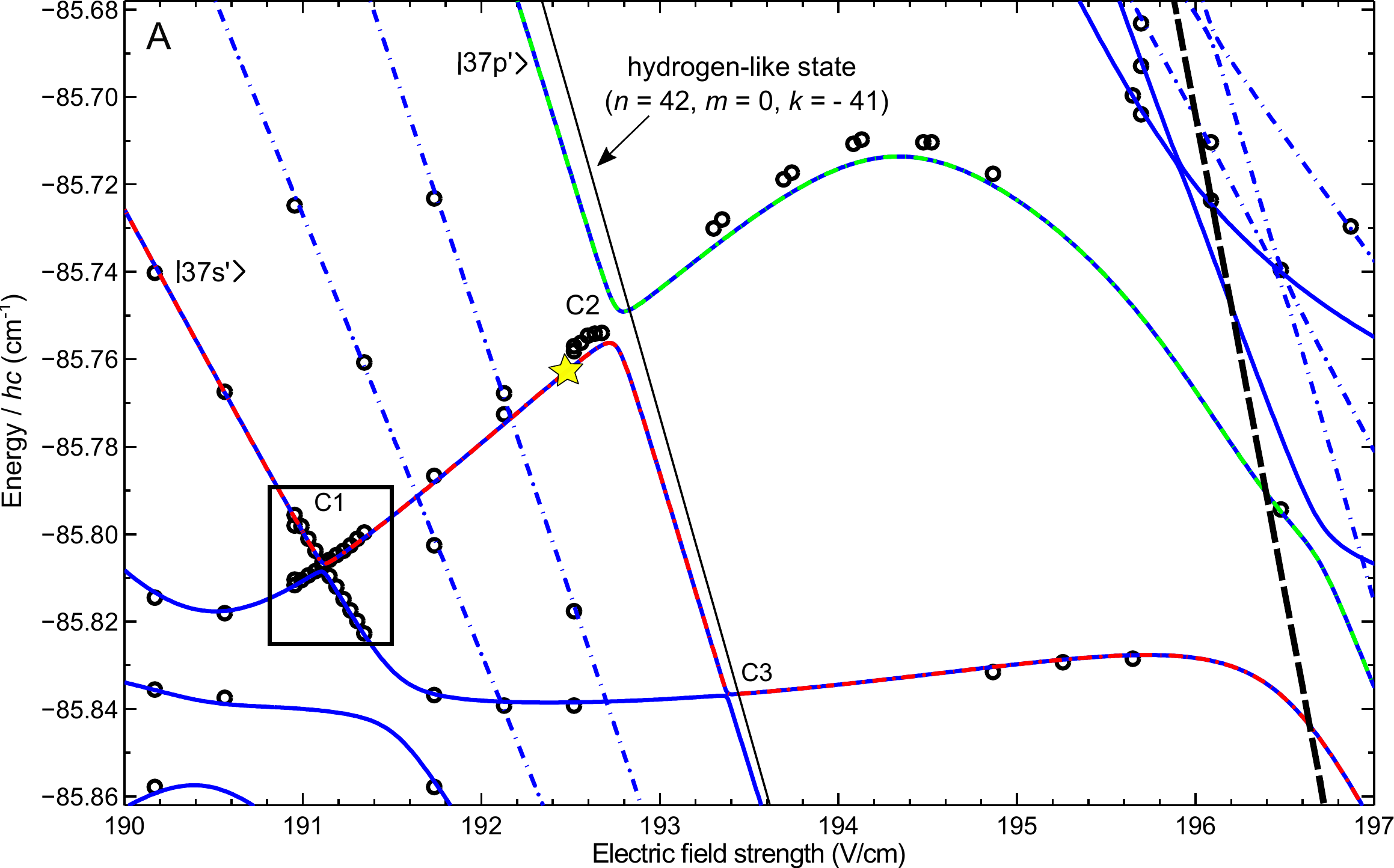}
\caption{The region of the Stark map enclosed in the solid rectangle labeled A in Fig.~\ref{fig1}. The calculated energy-level structure (thick continuous and dash-dotted curves) is compared to the positions of resonances in the measured laser spectra (open circles). The positions in the Stark map at which the $\big|$37s$'$$\big\rangle$ state was excited for the ionization rate measurements (see Sec. \Rmnum{4}) is marked by the yellow star. The thin continuous curve labeled $n$ = 42, $m$ = 0, $k$ = $-41$ represents to the hydrogen-like state discussed in the text.}\label{fig3}
\end{figure*}

As in Ref. \cite{Grimmel.2017}, to numerically calculate the eigenstate tunnel ionization rates, the total Hamiltonian for the system including the CAP is expressed as
\begin{equation}
\hat{H}_{\mathrm{CAP}} = \hat{H} - i\eta W,
\end{equation}
where $\eta$ is a scaling parameter that is unique to each eigenstate in each electric field of interest, and $W$ is the function chosen to act as the absorbing potential. In the experiments reported here, Rydberg state tunnel ionization rates within $\sim$ 1 cm$^{-1}$ of the Stark saddle point, and hence the classical ionization threshold, have been studied. Over this energy range the experimental data have been compared with the results of calculations performed using two different CAPs. Both of these potentials scale with $r^{6}$. In the first \cite{Sahoo.2000}
\begin{equation}
W(r) = r^6,
\end{equation}
while in the second \cite{Grimmel.2017}
\begin{equation}
W(r,f) = \Theta[r-r_\mathrm{c}(f)]\,[r-r_\mathrm{c}(f)]^{6},
\end{equation}
where $\Theta[r-r_\mathrm{c}(f)]$ is a Heaviside function, and $r_\mathrm{c}(f) = f^{-1/2}$ is the radial position associated with the Stark saddle point in the field $f$. The form of CAP in Eq. (4) has previously been used in the calculation of ionization rates of low-$n$ states in hydrogen and lithium in strong electric fields~\cite{Sahoo.2000}. The function in Eq. (5) corresponds to that employed in recent studies of Rydberg states in rubidium~\cite{Grimmel.2017}. This second form of CAP is a well defined function of the applied field and offers, in part, the possibility of removing the dependence of $\eta$ on the field strength. Because of the spherical symmetry of Eqs. (4) and (5), only the diagonal matrix elements associated with these CAPs, in the $|n\ell m\rangle$ basis in which the calculations were performed, were considered. The radial integrals required in the construction of these matrices were calculated using the Numerov method.\par 

The complex eigenvalues 
\begin{equation}
\lambda_j = E_j - i\frac{\Gamma_j}{2}
\end{equation}
of the Hamiltonian in Eq. (3), which includes the CAP, yield the eigenenergy $E_j$ and tunnel ionization rate $\Gamma_j$ of each eigenstate. The parameter $\eta$ in Eq. (3) can be obtained by comparison of $E_j$ with the energy of the corresponding eigenstate obtained from Eq. (2) in the absence of the CAP, or by comparison of $E_j$ and $\Gamma_j$ with experimental data. The values of the ionization rates, $\Gamma_j$, are more sensitive to changes in $\eta$ than the values of $E_{j}$. 

\section{\label{sec:level3}Experiment}
A schematic diagram of the experimental setup is shown in Fig.~\ref{fig2}(a). A pulsed supersonic beam of metastable helium atoms in the 1s2s $^{3}$S$_{1}$ level was generated in a dc electric discharge at the exit of a pulsed valve~\cite{Halfmann.2000}. The metastable atomic beam, traveling with a longitudinal speed of 2000 m/s, was collimated by a 2-mm-diameter skimmer and stray ions were filtered, before it entered the excitation region of the apparatus. The timing sequence for the excitation and ionization steps of the experiments is shown in Fig.~\ref{fig2}(b). Excitation of the atoms to Rydberg states with values of $n$ close to 37 took place at $t$ = 0 between two parallel copper electrodes labeled E$_1$ and E$_2$ in Fig.~\ref{fig2}(a). These electrodes were separated by 13~mm. The Rydberg states were prepared using a resonance-enhanced two-photon excitation scheme, with UV and IR laser beams \cite{Hogan.2018}. The UV laser was frequency stabilized to be resonant with the 1s2s~$^{3}$S$_{1}$ $\rightarrow$ 1s3p~$^3$P$_2$ transition at 25$\,$708.587$\,$6~cm$^{-1}$ ($\equiv$~388.975$\,$1 nm). The IR laser was tuned to subsequently drive 1s3p~$^3$P$_2$ $\rightarrow$ 1s$n$s/1s$n$d transitions. This excitation scheme was implemented in the presence of a strong electric field (190 -- 200~V/cm) generated by applying a high negative potential to electrode E$_2$. The laser radiation for both steps in this two-photon excitation scheme was linearly polarized parallel to this field. Consequently, the strong transitions in the spectra recorded were to states for which $|m| = 0$ and~1. 2.5 $\mu$s after photoexcitation, the electric field was switched for a short period of time, $T$, by applying a pulsed potential of between $-$10 and $+$10 V to electrode E$_1$ as indicated in Fig.~\ref{fig2}(b). This permitted the measurement of electric-field-dependent changes in the Rydberg state ionization rates. After the potential on electrode E$_1$ returned to its initial value, a slowly rising potential was subsequently applied at $t$ = 11.5 $\mu$s to electrode E$_2$ to ionize the surviving bound Rydberg atoms. The electrons resulting from this ionization step were accelerated to a microchannel plate (MCP) detector and collected for detection. In all of the measurements described here, the electron signal associated with the surviving Rydberg atoms was integrated and recorded to represent the number of detected atoms. 

\section{\label{sec:level4}Results}
The experiments described here were performed in the region of the Stark map in Fig.~\ref{fig1} between electric fields of 190 and 200 V/cm, and wavenumbers of $-86$ to $-85$~cm$^{-1}$. This part of the Stark map lies within 1~cm$^{-1}$, or 10~V/cm, of the classical adiabatic ionization thresholds of the $\big|$37s$'$$\big\rangle$ and $\big|$37p$'$$\big\rangle$  $\ell$-mixed eigenstates, which evolve adiabatically from the $\big|$37s$\big\rangle$ and $\big|$37p$\big\rangle$ Rydberg states in zero electric field, and is enclosed by the dashed rectangle labeled A in Fig.~\ref{fig1}.\par

\begin{figure}
	   \includegraphics[width=1\linewidth]{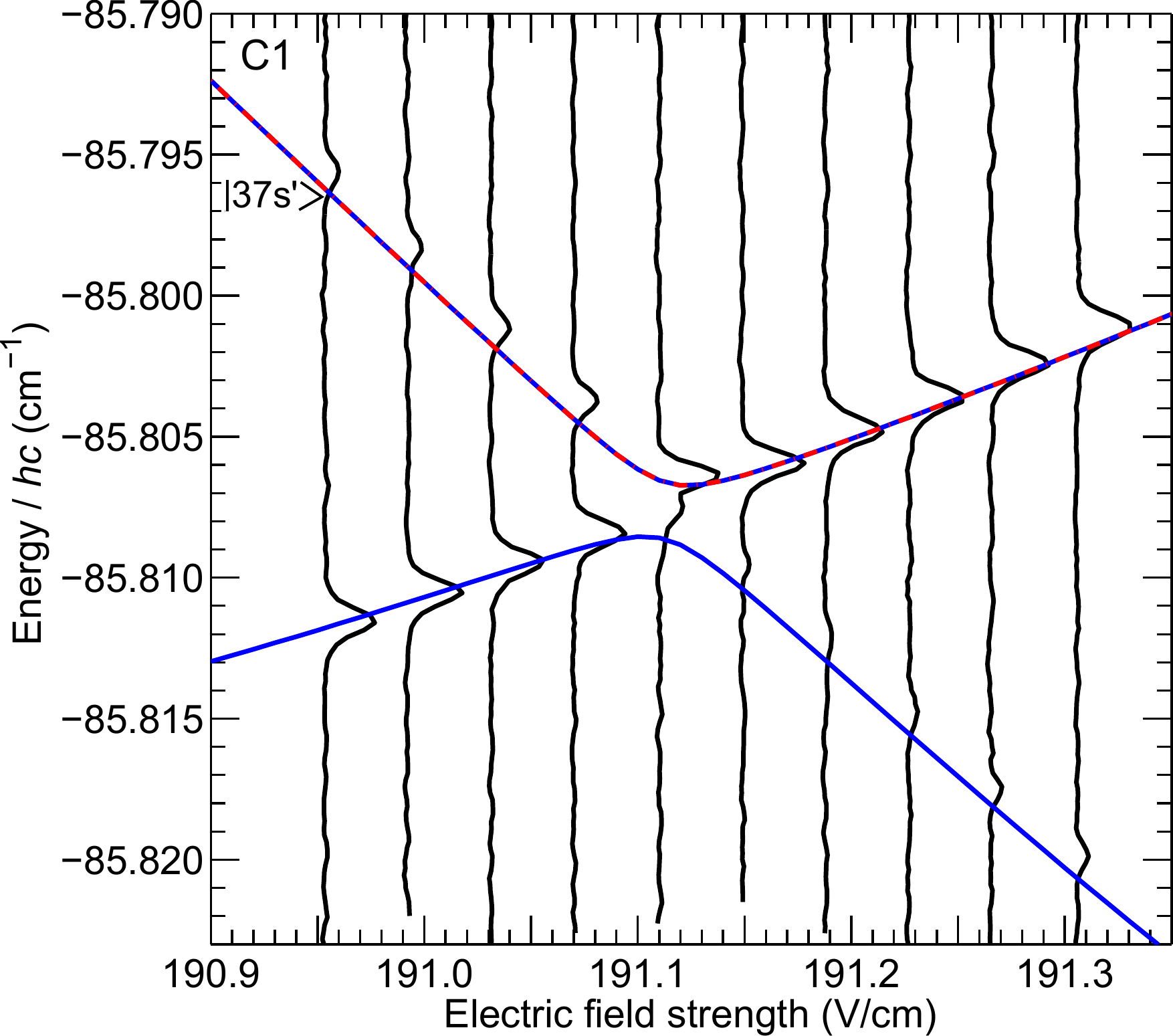}
\caption{Laser photoexcitation spectra (vertical curves) of triplet Rydberg states in helium overlaid with the calculated energy-level structure in the region of the avoided crossing labeled C1 in Fig.~\ref{fig3}. The amplitude of the spectral features corresponds to the integrated electron signal.}\label{fig4}
\end{figure}

An expanded view of this energetic region is displayed in Fig.~\ref{fig3}. In this figure the continuous and dashed-dotted curves correspond to the calculated energies of $m$ = 0 and $|m|$ = 1 sub-levels, respectively. The energy associated with the Stark saddle point, i.e., the classical ionization field, is indicated by the thick dashed black curve between 195.9 and 196.8 V/cm. To validate the results of the calculations, laser photoexcitation spectra of transitions to the eigenstates within the range of electric fields and energies encompassed in Fig.~\ref{fig3} were recorded. Examples of such spectra, in the region surrounding the avoided level crossing labeled C1, are displayed in Fig.~\ref{fig4}. In these data the observed spectral intensities arise from the combination of the relative transition strengths from the intermediate 1s3p\,$^{3}$P$_{2}$ level in each field, and excited state decay by tunnel ionization. The experimental spectra are compared with the calculated eigenenergies of the two states at the crossing, where the upper curve in Fig.~\ref{fig4} corresponds to the $\big|$37s$'$$\big\rangle$ state. There is very good quantitative agreement between the measured and calculated energy-level structure in these fields. Repeated measurements in this and other similar fields indicate that the deviation between the results of the experiments and calculations are dominated by slow fluctuations of $\sim\pm$~10~mV/cm (or $\pm$ 0.05\%) in the magnitude of the applied electric fields. The spectral widths of the measured transitions are limited by the interaction time of the atomic beam with the focused laser beams and are $\sim0.001$~cm$^{-1}$ \cite{Hogan.2018}. To obtain a broader perspective on the agreement between the experimental data and the results of the calculations, the wavenumbers associated with the the intensity maxima of each spectral feature in Fig.~\ref{fig4} were determined and overlaid, together with the results of similar measurements in a range of other electric fields, on the calculated energy-level diagram in Fig.~\ref{fig3}. Each open circle in this figure represents an observed transition to a Rydberg state with a sufficient lifetime ($\gtrsim 1~\mu s$) to permit detection. \par

\begin{figure}
	   \includegraphics[width=0.9\linewidth]{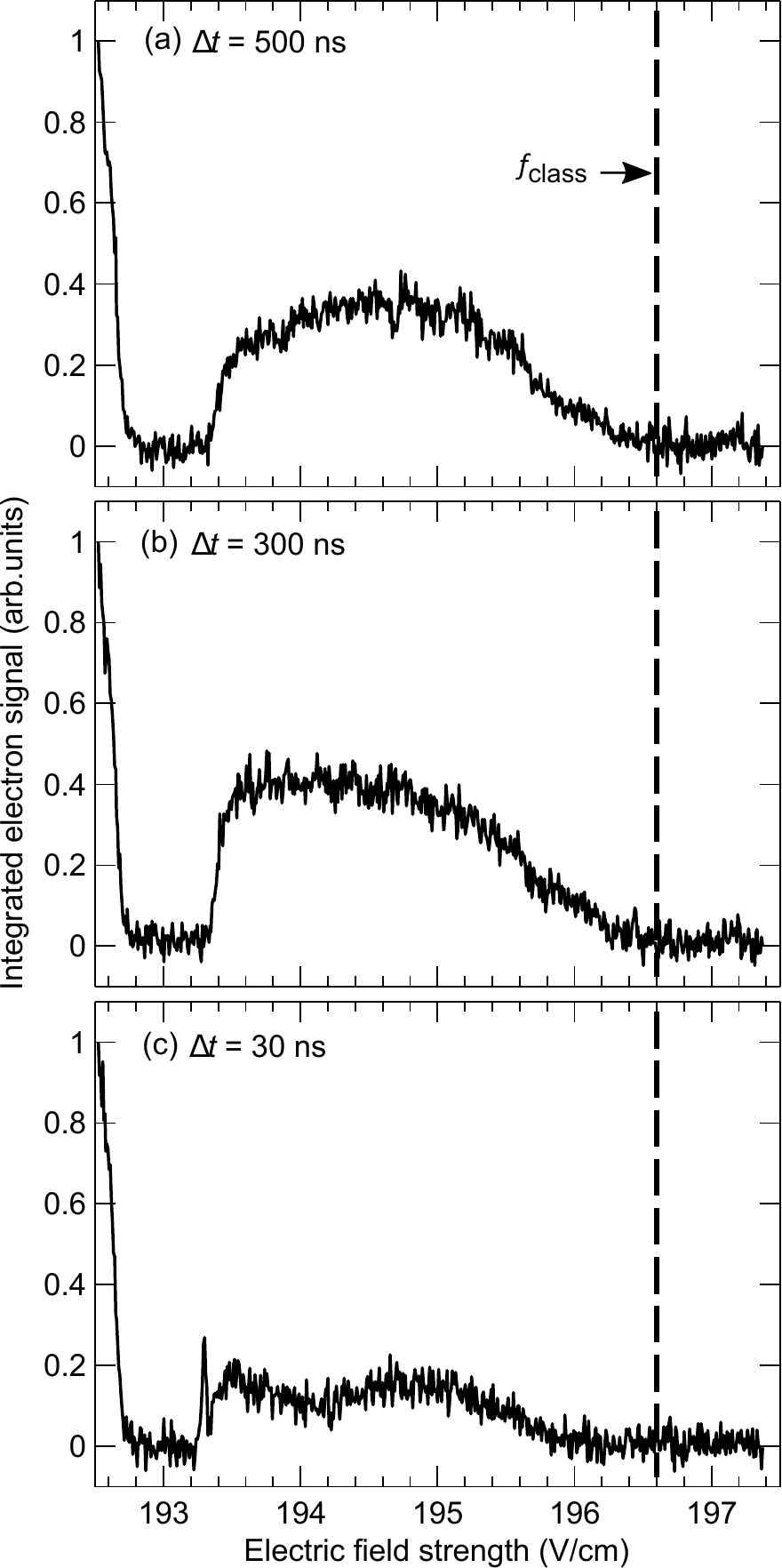}
\caption{The integrated electron signal recorded following laser photoexcitation of the triplet $\big|$37s$'$$\big\rangle$ state in helium, as a function of electric field strength for the three different field-switching times, $\Delta t$, as indicated.}\label{fig5}
\end{figure}

Rydberg state ionization dynamics, over the range of electric fields and energies encompassed in Fig.~\ref{fig3}, were studied by photoexcitation of each state of interest in a field in which its tunnel ionization rate was slow compared with the timescale of the experiment. The electric field was then switched to a higher or lower value to probe the field-dependence and time-dependence of the ionization dynamics. A first set of experiments were performed to study the threshold ionization fields and ionization rates of the $\big|$37s$'$$\big\rangle$ state. In these measurements, the atoms were excited to the $\big|$37s$'$$\big\rangle$ state in a field of 192.5 V/cm (yellow star in Fig.~\ref{fig3}) up to which the tunnel ionization rate is $<$ 10$^{5}$ s$^{-1}$. After photoexcitation the electric field in the apparatus was switched, in a time $\Delta t$, to a range of different values for $T$ = 4~$\mu$s [see Fig.~\ref{fig2}(b)]. After this, the field was switched back to its initial value and the fraction of surviving bound Rydberg atoms was determined. This procedure, performed for switching times $\Delta$$t$ = 500, 300 and 30 ns, resulted in the data presented in Fig.~\ref{fig5} and yielded insights into the ionization dynamics of the $\big|$37s$'$$\big\rangle$ state in this range of electric fields. \par 

As can be seen in Fig.~\ref{fig3}, the $\big|$37s$'$$\big\rangle$ state crosses the classical ionization threshold in a field of $\sim$ 196.6~V/cm. However, the data in Fig.~\ref{fig5} show that significant decay by ionization occurs in fields up to $\sim$ 4 V/cm below this threshold. When the field is switched slowly [$\Delta t$~=~500~ns, Fig.~\ref{fig5}(a)] from its initial value of 192.5~V/cm to values up to 197.4~V/cm, significant signal loss is seen in fields between 192.8 and 193.3 V/cm. Over this range of fields the $\big|$37s$'$$\big\rangle$ state evolves adiabatically through the avoided crossing C2 (see Fig.~\ref{fig3}) from being a low-field-seeking state, with a positive Stark shift, and hence electron density predominantly located on the side of the He$^+$ ion core opposite to that of the Stark-saddle point, to high-field-seeking with a negative Stark shift. In this case, the electron charge density is transferred to the side of the He$^+$ ion core close to the Stark saddle point, resulting in a higher propensity to tunnel ionize. Since in this range of fields almost all of the atoms are lost by ionization within the duration of the 4-$\mathrm{\mu}$s-long electric field pulse, the tunnel ionization rate must be in excess of $\sim$ 2.5 x 10$^{5}$ s$^{-1}$. For fields higher than 193.3~V/cm the $\big|$37s$'$$\big\rangle$ state passes through the avoided crossing C3 in Fig.~\ref{fig3}. The linear Landau-Zener model~\cite{Landau.1932,Zener.1932} indicates that when $\Delta t$ = 500 ns (d$F$/d$t$~$\sim$~2 V/cm/$\mu$s) the probability of adiabatic traversal of this avoided crossing is $\sim$~1. This is consistent with the increase in signal in Fig.~\ref{fig5}(a) for fields between 193.4 and 196 V/cm that occurs because the $\big|$37s$'$$\big\rangle$ state switches from being strongly high-field-seeking to having approximately zero static electric dipole moment. Therefore, there is an almost equal distribution of electron charge on both sides of the He$^+$ ion core, close to and far from the Stark saddle point. This results in a reduced propensity to tunnel ionize. Finally, as the field approaches the classical ionization field of 196.6~V/cm the loss of excited atoms increases. This increase arises in part because of the depression of the Stark saddle point in these fields, and in part because the $\big|$37s$'$$\big\rangle$ state becomes slightly more polarized, and high-field-seeking, in fields above 196~V/cm (see Fig.~\ref{fig3}). \par

\begin{figure}
\includegraphics[width=1\linewidth]{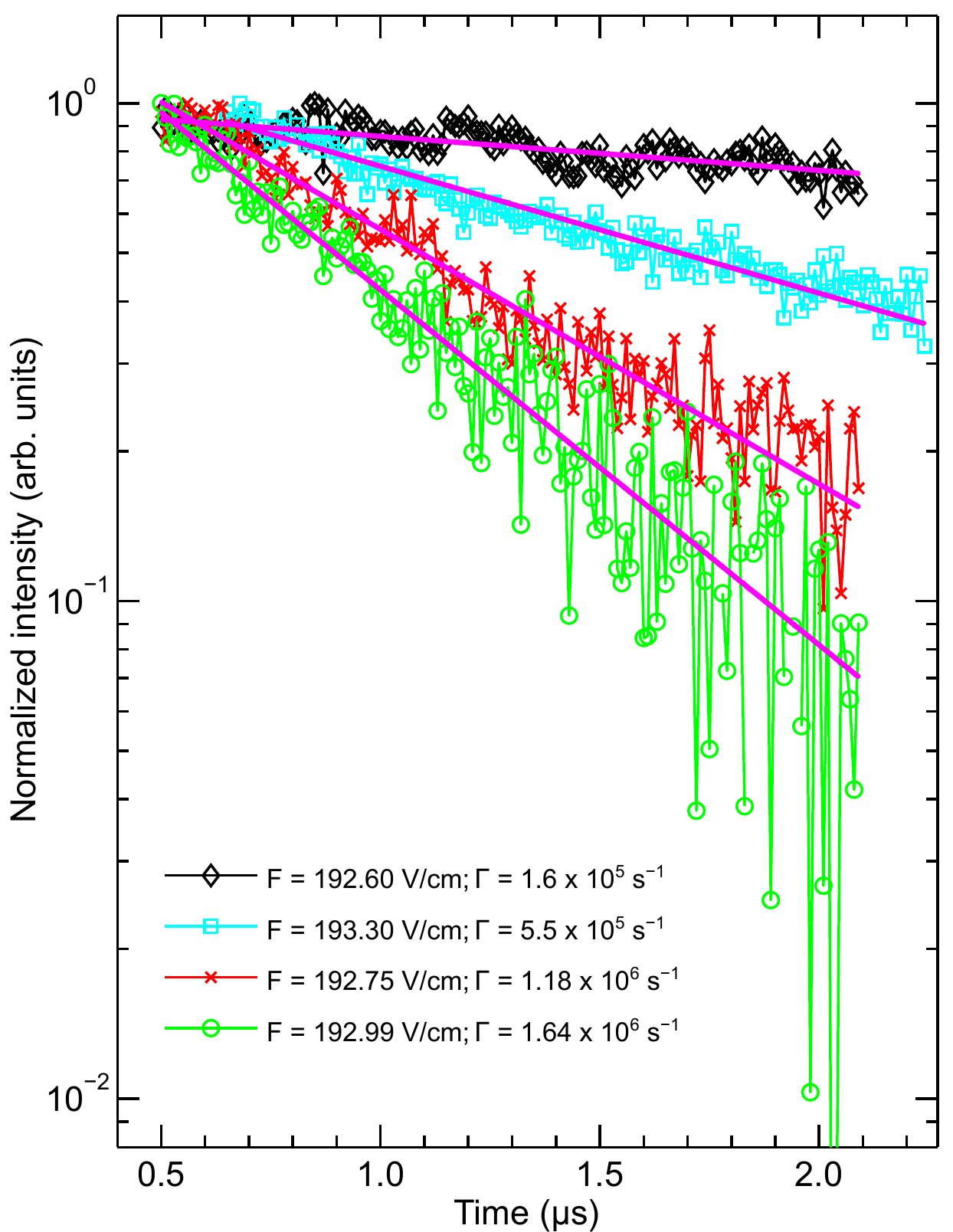}
\caption{Tunnel ionization of the triplet $\big|$37s$'$$\big\rangle$ state in helium recorded in the electric fields indicated. Fitted exponential functions (straight lines) yielded the ionization rates listed.}\label{fig6}
\end{figure}

A comparable set of measurements carried out for $\Delta t$~=~300~ns [Fig.~\ref{fig5}(b)] yielded similar results. The only notable difference between these two sets of data is the slightly larger fraction of surviving Rydberg atoms in fields between 193.4 and 194.8 V/cm when $\Delta t$~=~300~ns. This is a result of switching more quickly through the region between 192.8 and 193.2 V/cm in which rapid tunnel ionization occurs. Therefore, the ionization dynamics in Fig.~\ref{fig5}(a) and Fig.~\ref{fig5}(b) must be considered similar, with a similar evolution through the avoided crossings C2 and C3. However, if $\Delta t$ is further reduced to 30 ns (d$F$/d$t$~$\sim$ 30 V/cm/$\mu$s) the surviving Rydberg atom fraction in fields beyond 193.2 V/cm is quite different [see Fig.~\ref{fig5}(c)]. In this case, the avoided crossing C2 is traversed adiabatically, while the Landau-Zener probability of adiabatic traversal of crossing C3 reduces to $\sim$ 0.6. After passing through the crossing C3, the excited atoms evolve into a superposition of the two interacting states with their significantly different Stark energy shifts and hence tunnel ionization rates. The time-evolution of this superposition contributes to the oscillatory behavior of the surviving fraction of Rydberg atoms in fields between 193.2 and 196 V/cm. \par

In experiments in which ramped electric field ionization of the $\big|$37s$'$$\big\rangle$ state is implemented beginning in fields close to zero, the electric field switching rates are typically on the order of 200 V/cm/$\mu$s \cite{Zhelyazkova.2017}. Consequently, under such conditions the turning point at the avoided crossing C2 at 192.8 V/cm may be considered the threshold adiabatic ionization field of the $\big|$37s$'$$\big\rangle$ state. \par

More detailed information can be obtained on the tunnel ionization rates of the states in Fig.~\ref{fig3} by direct measurements in the time domain. To perform these for the $\big|$37s$'$$\big\rangle$ state, laser photoexcitation was again carried out at the same position in the Stark map as when recording the data in Fig.~\ref{fig5} (yellow star in Fig.~\ref{fig3}). However, after this was completed, the probe electric field pulse ($\Delta t$~=~300~ns) was applied for period of time, $T$, that was adjusted in the range from 500 ns to 2 $\mu$s, The field was then switched back to its initial value of 192.5~V/cm before the surviving fraction of excited atoms were detected by pulsed electric field ionization. The dependence of the integrated electron signal on the probe pulse duration was then recorded for a range of probe fields, with the corresponding loss rate of excited atoms reflecting the tunnel ionization rate in each field. Examples of this data, recorded for pulsed electric fields of 192.60, 192.75, 192.99, and 193.30 V/cm, are displayed in Fig.~\ref{fig6}. The measured loss rates in each case can be described by a single exponential function. For the range of fields studied the time constants of these functions range from 600~ns to 6.25~$\mu$s. \par

\begin{figure}
\centering
	   \includegraphics[width=1\linewidth]{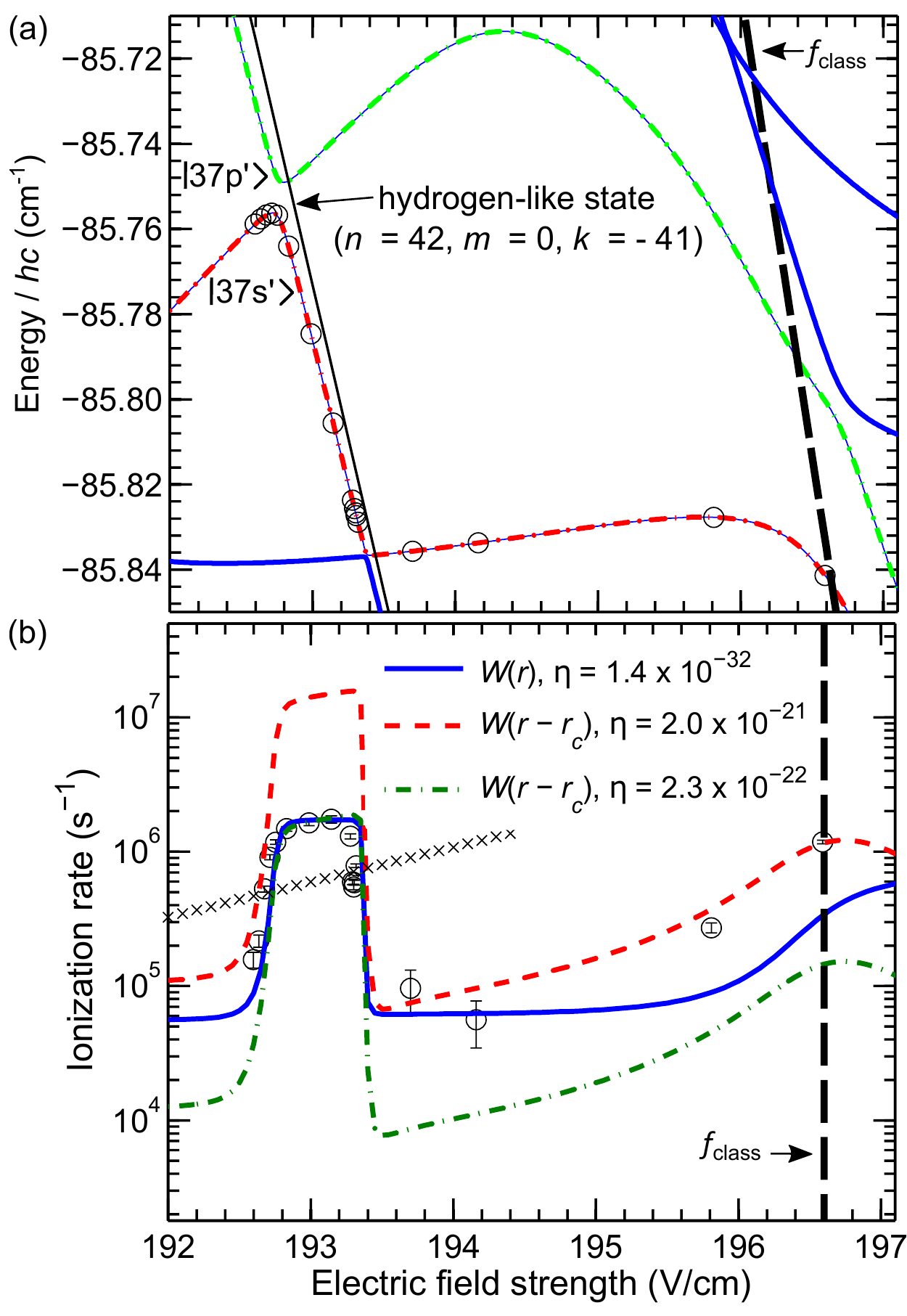}
\caption{(a) The fields and energetic positions at which the ionization rates of the $\big|$37s$'$$\big\rangle$ state were measured (open circles). (b) The corresponding ionization rates (open circles) together with the results of the calculations. Ionization rates calculated using the field-independent CAP are shown as the continuous blue curve in (b), while those calculated with the field-dependent CAP are represented by the dashed and dashed-dotted curves for the two different values of $\eta$ listed.}\label{fig7}
\end{figure}

A complete set of measured tunnel ionization rates of the $\big|$37s$'$$\big\rangle$ state, determined from data recorded in pulse electric fields between 192.6 and 196.59 V/cm, are presented in Fig.~\ref{fig7}. To allow direct comparison with the corresponding energies of the $\big|$37s$'$$\big\rangle$ state a portion of the Stark map in Fig.~\ref{fig3}, containing only the states for which $m$ = 0, is included in Fig.~\ref{fig7}(a). The positions at which the ionization rates were measured are shown as open circles in this Stark map, with the corresponding ionization rates also indicated as open circles in Fig.~\ref{fig7}(b). \par

As can be seen from Fig.~\ref{fig7}(a) and the experimental data in Fig.~\ref{fig7}(b), in fields between 192.6 and 192.8 V/cm the $\big|$37s$'$$\big\rangle$ state is low-field-seeking and its tunnel ionization, measured to occur at a rate between 1.6 x 10$^5$ and 5.3 x 10$^5$ s$^{-1}$, is suppressed because the electron charge is located predominantly on the side of the He$^+$ ion core opposite to the Stark saddle point. When it becomes high-field-seeking, in fields between 192.8 and 193.3~V/cm, the tunnel ionization rate immediately increases to $>$~10$^6$~s$^{-1}$. This change is a result of the electron charge being transferred to the side of the He$^+$ ion core close to the Stark saddle point, and occurs despite the fact that in this range of fields the $\big|$37s$'$$\big\rangle$ state is more deeply bound than in the lower fields of 192.6 to 192.8~V/cm. 

The experimental data in Fig.~\ref{fig7}(b) have been compared to the results of calculations of the corresponding tunnel ionization rates performed using the CAP methods described in Sec. \Rmnum{2}. The results of the calculations in which the electric-field-dependent form of the CAP in Eq.~(5) was used are indicated by the dashed and dash-dotted curves. The use of the field-independent CAP in Eq.~(4)  yielded the results indicated by the continuous curve. In using both of these CAPs, the value of the scaling factor $\eta$ was obtained, in atomic units, by fitting the calculated results to the experimental data. \par

For the electric-field-dependent CAP, the value of $\eta$ was first fitted to the measured ionization rates in excess of 10$^6$ s$^{-1}$ in fields between 192.7 and 193.2 V/cm. This resulted in a value of $\eta$ = 2.3 x 10$^{-22}$ and the data indicated by the dash-dotted curve in Fig.~\ref{fig7}(b). The results of the calculations are in good quantitative agreement with the experimental data over the range of fields in which high ionization rates were measured, i.e., when the $\big|$37s$'$$\big\rangle$ is high-field-seeking. But when the static electric dipole moment of the $\big|$37s$'$$\big\rangle$ state reduces towards zero, in fields above 193.5 V/cm, the calculated rates are approximately an order of magnitude lower than those measured in the experiments. With the measured ionization rates extending over almost two orders of magnitude in the range of fields in Fig.~\ref{fig7}(b) the results of the calculations do however follow the same general trend as the field strength increases. \par

In fields above 193.5 V/cm, where the results obtained with $\eta$ = 2.3 x 10$^{-22}$ deviate most significantly from the experimental data, it was found that calculations in which $\eta$~=~2.0 x 10$^{-21}$ [dashed curve in Fig.~\ref{fig7}(b)] yielded closer agreement with the experimental data. However, in this case the calculated rates deviate from the experimental data by a factor of approximately 10 in the low-field region between 192.7 and 193.2 V/cm. These observations indicate that using the electric-field-dependent form of the CAP in Eq. (5) is not sufficient to completely account for effects of the sign of the static electric dipole moments, and hence Stark shifts, of the states, on the tunnel ionization rates below the classical ionization threshold. In these instances it would seem that it may be necessary to combine the electric-field-dependent form of $W$($r,f$) with a dependence of $\eta$ on the static electric dipole moment, or derivative of the energy with respect to the field strength, of the states of interest to obtain more reliable results over a wider range of fields. \par

Further comparison of the experimental data in Fig.~\ref{fig7}(b) with the results of calculations performed with the electric-field-independent CAP in Eq. (4) was also carried out. In this case the most appropriate value of $\eta$ to fit the results of the calculations to the experimental data in the low-field region between 192.7 and 193.8 V/cm was $\eta$ = 1.4 x 10$^{-32}$ [continuous blue curve in Fig.~\ref{fig7}(b)]. This value of $\eta$ gives good quantitative agreement with the experimental data over a wider range of fields than the calculated results obtained using the field-dependent CAP. Significant deviations are only seen when the energy of the $\big|$37s$'$$\big\rangle$ state approaches the classical ionization threshold. \par

\begin{figure}
\centering
	   \includegraphics[width=1\linewidth]{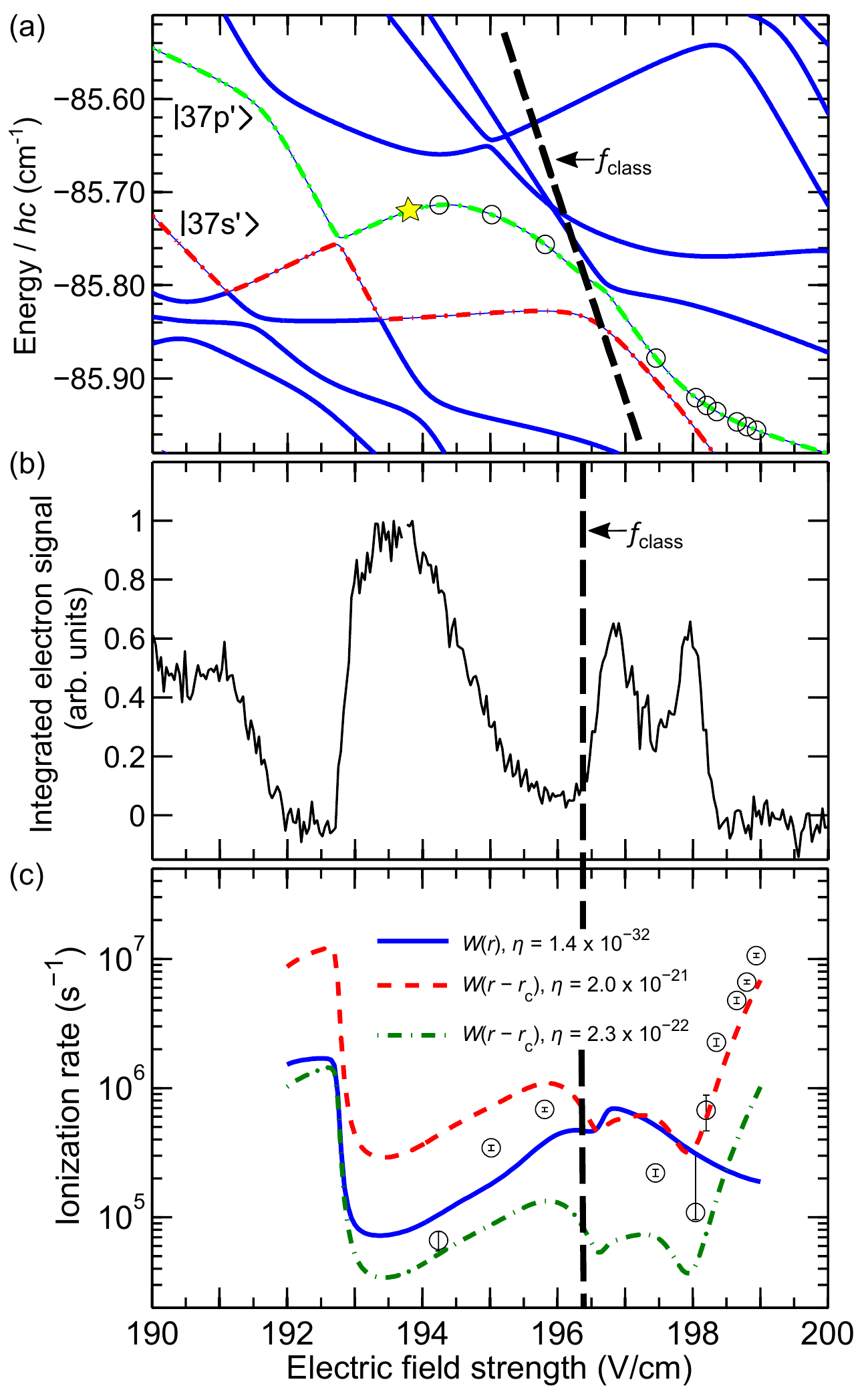}
\caption{(a) The fields at which the ionization rates of the $\big|$37p$'$$\big\rangle$ state were measured (open circles) and the initial photoexcitation field (yellow star). (b) The integrated electron signal of the  $\big|$37p$'$$\big\rangle$ state for $T = 4~\mu$s and $\Delta t=300$~ns. (c) Measured (open circles) and calculated tunnel ionization rates. The ionization rates calculated using the field-independent CAP are indicated by the continuous blue curve in (c), while those calculated with the field-dependent CAP are represented by the dashed and dashed-dotted curves for the two different values of $\eta$ indicated.}\label{fig8}
\end{figure}

Finally, the ionization rates measured for the $\big|$37s$'$$\big\rangle$ state can be compared with the tunnel ionization rates for hydrogenic Rydberg states calculated using the semi-empirical approach in Ref. \cite{Damburg.1979}. The hydrogen-like state with $n$ = 44, $m$ = 0, $k$~=~$n_1 - n_2$ = $-$41, where $n_1$ and $ n_2$ are the parabolic quantum numbers [thin black line between 192.4 and 193.6~V/cm in Fig.~\ref{fig3} and Fig.~\ref{fig7}(a)] lies within $\sim$ 0.01 cm$^{-1}$ of the energy of the $\big|$37s$'$$\big\rangle$ state in fields between 192.3 and 193.8~V/cm. These are the fields that correspond to the highest measured ionization rates in Fig.~\ref{fig7}(b). The calculated ionization rates for this hydrogen-like state are indicated by the crosses in Fig.~\ref{fig7}(b). The semi-empirical calculations for this hydrogen-like state yield tunnel ionization rates that are a factor of 5 lower than those measured for the $\big|$37s$'$$\big\rangle$ state. This observation gives an indication of the accuracy of this semi-empirical method widely used in the calculation of tunnel ionization rates of hydrogenic Rydberg states~\cite{Alonso.2018}, and is of a similar scale to the expected deviations from the exact rates discussed in Ref.~\cite{Damburg.1979}. Indeed, because the $\big|$37s$'$$\big\rangle$ state is more tightly bound than the associated hydrogen-like state, it is likely that the semi-empirical approach underestimates the true ionization rates by more than the observed factor of 5. \par

The procedure outlined above to measure the ionization rates of the $\big|$37s$'$$\big\rangle$ state was also applied to determine the tunnel ionization rates of the $\big|$37p$'$$\big\rangle$ state. The position in the Stark map at which this state was initially laser photoexcited in these experiments is indicated by the yellow star in Fig.~\ref{fig8}(a) and the fields in which ionization rates were measured are shown as open circles. An overview of the field-dependence of the tunnel ionization of the $\big|$37p$'$$\big\rangle$ state was first obtained by subjecting the excited atoms to a range of pulsed electric fields for $T= 4~\mu$s with $\Delta t=300$~ns. The results of these measurements, which indicate the fraction of atoms surviving in bound states following this process are displayed in Fig.~\ref{fig8}(b). These data indicate two regions of significant tunnel ionization below the classical ionization field: 192.0 to 192.8~V/cm and 195.8 to 196.2~V/cm. As in the case of the $\big|$37s$'$$\big\rangle$ state, these are regions in which the $\big|$37p$'$$\big\rangle$ state exhibits pronounced high-field-seeking character. The ionization rates measured at the fields indicated in Fig.~\ref{fig8}(a) are shown as open circles in Fig.~\ref{fig8}(c), together with the rates calculated using the field-independent (continuous curve) and field-dependent (dashed and dash-dotted curves) CAPs. The values of $\eta$ in these CAPs are the same as those used in the calculations for the ionization rates of the $\big|$37s$'$$\big\rangle$ state. Below the classical ionization field the results obtained using the field-independent CAP are in good agreement with the experimental data. In fields above $f_{\mathrm{class}}$ the field-dependent CAP with $\eta$~=~2.0 x 19$^{-21}$ provides the closest agreement with the experimental data. But, as in the case of the $\big|$37s$'$$\big\rangle$ state, no one CAP completely describes the ionization dynamics over the full range of fields in Fig.~\ref{fig8}. These observations for the $\big|$37p$'$$\big\rangle$ state lead to the same conclusion as that reached for the $\big|$37s$'$$\big\rangle$ state from the data in  Fig.~\ref{fig7}(b): In order to calculate ionization rates that accurately reflect the measured values, a field-dependent CAP with an additional dependence on the static electric dipole moment of the states of interest is needed. \par

It is evident from the data in Fig.~\ref{fig8} that the measured electron signal of the $\big|$37p$'$$\big\rangle$ state is dependent on whether it is high-field-seeking or low-field seeking. Comparing the data in Fig.~\ref{fig8}(b) with the ionization rates calculated using the field-dependent CAPs in Fig.~\ref{fig8}(c) indicates that when the measured electron signal is low, i.e., the lifetime of the Rydberg atoms is not long enough to permit detection, the corresponding calculated ionization rate is higher than the ionization rate in a field where the electron signal is high. In general, the tunnel ionization rates calculated using the field-dependent CAPs match each corresponding structure in Fig.~\ref{fig8}(b). For example, the two intensity maxima in Fig.~\ref{fig8}(b), in fields between 196.4 and 198.4 V/cm correspond to two dips in the ionization rate in Fig.~\ref{fig8}(c) in the same range of fields. It can also be seen in Fig.~\ref{fig8}(b) that there remain structures observed beyond the classical ionization threshold at 196.4~V/cm for the $\big|$37p$'$$\big\rangle$ state. However, in fields from 198 to 208~V/cm the electron signal does not rise above zero, suggesting that this state always ionizes rapidly in fields higher than the classical ionization field. Indeed, at 199 V/cm the measured ionization rate is on the order of 10$^{7}$ s$^{-1}$ and for fields slightly lower than this the signal in Fig.~\ref{fig8}(b) is already zero. \par

\section{\label{sec:level5}Conclusions}
In conclusion, ionization dynamics and tunnel ionization rates of triplet Rydberg states in helium have been studied in fields within a factor of 0.01 of the classical ionization electric field. The experimental data, including direct measurements of tunneling rates in the time domain, have been compared to the results of calculations performed using two types of complex absorbing potential. The calculated energy-level structure is in excellent quantitative agreement with the experimental data over the full range of fields and eigenstates studied. The calculated tunnel ionization rates follow the same general trends as those seen in the experiments. However, the precision of the time-domain measurements, which encompass almost two orders of magnitude in tunneling rate, highlights deficiencies in these methods of calculation. They also exceed the limit of the applicability of tunnel ionization rate calculations for hydrogen-like Rydberg states with similar characteristics. \par
The results presented are of importance in the optimization of state-selective electric field ionization schemes employed in Rydberg-atom$-$polar-molecule collision studies \cite{Zhelyazkova.2017,Zhelyazkova.2017b}. They are also of interest for experiments in which helium atoms in low-$\ell$ Rydberg states are used to study charge transfer processes in collisions with surfaces \cite{Gibbard.2015}. The general observations made suggest that the Rydberg states studied here can act as a valuable testing ground for calculations of tunnel ionization in strong electric fields. They offer opportunities to address aspects of symmetry breaking in tunnel ionization in perpendicular electric and magnetic fields~\cite{Zhang.2018}, and effects of the tunneling time in electric field ionization dynamics~\cite{Hauge.1989}.

\begin{acknowledgments}
This work is supported by the European Research Council (ERC) under the European Union's Horizon 2020 research and innovation program (Grant No. 683341).
\end{acknowledgments}

\bibliography{references}

\end{document}